\documentclass[11pt]{article}
\usepackage[T1]{fontenc}
\usepackage[utf8]{inputenc}
\usepackage{lmodern}
\usepackage[a4paper,margin=27mm]{geometry}
\usepackage{amsmath,amssymb,amsthm,mathtools}
\usepackage{enumitem}
\usepackage{microtype}
\usepackage{tikz}
\usepackage[hidelinks]{hyperref}
\usepackage[capitalise,noabbrev]{cleveref}

\usepackage{authblk}
\newtheorem{theorem}{Theorem}[section]
\newtheorem{lemma}[theorem]{Lemma}
\newcommand{\OPT}{\operatorname{OPT}}
\newcommand{\len}{\operatorname{len}}
\newcommand{\R}{\mathbb R}
\newcommand{\E}{\mathbb E}
\newcommand{\eps}{\varepsilon}
\newcommand{\cP}{\mathcal P_k}
\title{An efficient polynomial-time approximation scheme for\\the weighted Euclidean Minimum Latency Problem\\in fixed dimension}
\author[1]{Daniel Skachkov\thanks{Corresponding author.
  Email: \texttt{skach370@gmail.com}}}
\author[1]{Yuriy Dorn}
\affil[1]{MSU Institute for Artificial Intelligence, Moscow, Russia}
\date{}
\begin{document}

\maketitle

\begin{abstract}
In the weighted Minimum Latency Problem, also known as the Traveling Repairman Problem, a route starts at a depot and visits all customers. The objective is to minimize the weighted sum of the times at which customers are first visited. We give a deterministic efficient polynomial-time
approximation scheme (EPTAS) for the weighted Euclidean Minimum Latency
Problem in every fixed dimension $d$. For every $\varepsilon>0$, the algorithm returns a route of
cost at most $(1+\varepsilon)$ times the optimum using
$f_d(\varepsilon)\mathcal{B}^{O_d(1)}$ bit operations, where
$\mathcal{B}$ is the total binary encoding length of the input.
For every fixed $d\ge2$, we also rule out a fully polynomial-time
approximation scheme (FPTAS) unless $\mathrm{P}=\mathrm{NP}$,
even for unit weights and integer coordinates.
\end{abstract}


\section{Introduction}\label{sec:intro}

In the weighted Minimum Latency Problem (MLP), also known as the
Traveling Repairman Problem (TRP), a route starts at a root
and visits a set of customers.
The objective is to minimize $\sum_v w_vC_v$, where $w_v$ is the given customer's weight and $C_v$ the time of its first visit. In the Euclidean version, travel time is Euclidean distance.
Each edge contributes to the latency objective its length
multiplied by the total weight of customers still waiting
for service. Thus a short route can have large weighted latency.

Arora and Karakostas~\cite{AK2003} gave a quasi-polynomial-time
approximation scheme for Euclidean MLP in fixed dimension.
Sitters~\cite{Sitters2021} gave a polynomial-time approximation scheme
for weighted MLP in the plane, on trees, and on planar graphs.
He splits a route into phases and shows
that within each phase a near-optimal route can be divided into a
constant number of blocks. We use this decomposition. Sitters' planar
algorithm runs in $n^{\operatorname{poly}(1/\eps)}$ time. By rounding
the weight assigned to each block within each dissection cell, we obtain
an exponent of $n$ that depends only on the dimension.
Our main result is described by the following theorem.

\begin{theorem}\label{thm:main}
For every fixed positive integer $d$, weighted Euclidean
MLP in $\R^d$ admits a deterministic EPTAS. For every $\eps>0$,
it returns a route of cost at most $(1+\eps)\OPT$ using
$f_d(\eps)(n+1)^{O_d(1)}$ arithmetic operations, where
$f_d(\eps)$ is independent of the number $n$ of customers.
\end{theorem}

For rational input of total binary encoding length $\mathcal B$,
the bit running time is
\[
f_d(\eps)(n+\mathcal B+1)^{O_d(1)}.
\]
Thus the polynomial exponent is independent of $\eps$ in both models.

In \cref{sec:no-fptas}, we also give a short
gap argument showing that, for every fixed $d\ge2$, no fully polynomial-time approximation scheme (FPTAS) exists
unless $\mathrm P=\mathrm{NP}$, even with unit customer weights
and distinct integer locations. This is a direct consequence of
the classical NP-completeness of Hamilton path in grid
graphs~\cite{IPS1982}. Dimension one is exceptional: the interval
dynamic program of Afrati et al.~\cite{Afrati1986} extends to
arbitrary nonnegative customer weights and solves the problem
exactly in polynomial time. We explain this extension in \cref{sec:one-dim}.

\subsection{Proof overview: three transformations}

We gradually restrict the form of the route until the remaining
optimization problem can be solved by dynamic programming. Each of the
following restrictions loses only a factor close to one.

\paragraph{1. Divide the route into phases.}
We first compute a constant-factor approximate route. We then
consider a finite family of partitions of time into intervals.
For each partition, customers whose completion times on this
route fall in the same interval form a phase. The route for each
phase starts at the root at a prescribed time and returns there
in time for the next phase. Averaging over the partitions shows
that at least one of them admits a route of nearly optimal cost.

\paragraph{2. Divide each phase into a constant number of blocks.}
Within each phase, we group consecutive visits into a constant
number of blocks and charge each customer at the end of its
assigned block. We show that the blocks can be chosen so that
this increases the cost by only a small factor. Once the total
weight assigned to each block is fixed, the cost is a weighted
sum of block lengths plus a fixed term for the phase's start time.

\paragraph{3. Restrict block crossings to portals.}
We add an auxiliary coordinate, called height, and place all
customers at height zero. For each block, we place a point at
height $H$ above each endpoint. Using the norm
$\|(x,z)\|_*=\|x\|_2+|z|$, we close the block into a tour through
these two points. We use Arora's argument~\cite{Arora1998} to
show that, with only a small increase in cost, the tours can be
made to cross the faces of a common dissection only at designated
points, called portals, and at most a bounded number of times
per face. From each tour we extract an arc that visits all
customers assigned to its block. We project these arcs back to
$\R^d$ and join them in block order to obtain a route for the phase.

\

\cref{sec:outer,sec:blocks,sec:geometry} prove the three
structural restrictions. \cref{sec:rounding,sec:dp}
give the algorithm for one phase, and \cref{sec:complete}
assembles the phases.

\section{Model and a completion operation}\label{sec:model}

The input consists of a root $r\in\R^d$ and a set $V$ of $n$
distinguishable customers with locations in $\R^d$ and weights $w_v\ge0$.
Different customers may have the same location. Coordinates and weights are rational numbers encoded in binary.
Travel speed is one, service takes no time, and all customers are
available at time zero. We describe the geometric constructions
using exact distances. \cref{app:arithmetic} gives their
implementation with rational arithmetic.
Write $d(u,v)=\|u-v\|_2$ and $W=\sum_v w_v$.
For a customer permutation $\pi$, with $\pi_0=r$, define
\[
 C_{\pi_j}=\sum_{h=1}^j d(\pi_{h-1},\pi_h),\qquad
 F(\pi)=\sum_{j=1}^n w_{\pi_j}C_{\pi_j}.
\]
Here $C_{\pi_j}$ is the arrival time at the listed occurrence of
$\pi_j$. Let $\OPT$ be the minimum of $F(\pi)$ over all permutations.
Listing customers in order of their first visits on any walk and
shortcutting gives a permutation of no greater cost. Thus walks and
permutations have the same optimum. Intermediate constructions may
use repeated visits, waiting, and auxiliary destinations.

Serve customers at the root at time zero. Set aside zero-weight
customers and append any still unvisited ones at the end. These
operations do not change the optimum. If no positive-weight customers
remain away from the root, the problem is solved. Henceforth all
customers under consideration have positive weight.

A \emph{delayed instance} starts at the root at time $L>0$, with
\begin{equation}\label{eq:delayed}
 F_L(\pi)=WL+F(\pi),\qquad
 \OPT_L=WL+\OPT.
\end{equation}
It is \emph{$\delta$-bounded} if a known path from the root visits
all its customers and has length at most $\delta L$.
For a delayed instance, completion times include the delay $L$.

The following operation has two purposes: to make consecutive phases fit together, and to bound the time horizon within a phase.

\begin{lemma}[{Adapted from Sitters~\cite[Lemma~2.3]{Sitters2021}}]\label[lemma]{lem:complete}
Fix a delayed instance with customer set $S$, root $r$, and delay $L$.
Suppose a known path $\mathcal H$ starts at $r$, visits every customer in $S$,
and has length at most $D$. For $0<\xi\le1$, any route $\mathcal P$ for this
instance with delayed cost $Z$ can be changed into a closed tour of length at most $5D/\xi$ with delayed cost that does not exceed $(1+\xi)Z$.
\end{lemma}
\begin{proof}
Every customer lies within distance $D$ of $r$, by the bound on
the length of $\mathcal H$. After shortcutting the given route
$\mathcal P$, it is also within distance $D$ of $r$.

Set $T=2D/\xi$ and retain the longest prefix of the shortcut route
whose travel length is at most $T$. Follow that prefix, return
to $r$, follow $\mathcal H$, and return to $r$ again. The prefix
may be empty or may contain all customers. The endpoint of
$\mathcal H$ is also within distance $D$ of $r$, so each return
has length at most $D$. Thus the resulting closed route has
length at most
\[
 T+3D=\frac{2D}{\xi}+3D\le\frac{5D}{\xi}.
\]

Every customer visited in the retained prefix is served no later
than before. Every other customer had completion time $C>L+T$
on the shortcut route and now has completion time at most
$L+T+2D$. Since $2D=\xi T$,
\[
 L+T+2D\le(1+\xi)(L+T)<(1+\xi)C.
\]
Multiplying these bounds by the nonnegative customer weights
and summing proves the cost guarantee.
\end{proof}

\subsection{One-dimensional case} \label{sec:one-dim}

On a line, weighted MLP can be solved exactly by adapting the
interval dynamic program of Afrati et al.~\cite{Afrati1986}.
We merge customers at the same location by summing their weights,
then sort the distinct customer locations together with the root.

A state records the interval of visited locations and the endpoint
at which the server is located. Initially, the interval consists
only of the root. The next first visit must be to the nearest
unvisited location on either side of the interval, since reaching
any farther location requires passing through it. The server moves
directly to the chosen location, which becomes the new endpoint
of the interval.

The cost of a transition is the distance traveled multiplied by
the total weight of customers outside the current interval. This
is exactly the weighted latency accumulated during the move,
since those customers are still waiting. Prefix sums give the
total weight outside the interval in constant time.

A terminal state covers all customer locations. There are
$O(n^2)$ states and at most two transitions per state, so the
dynamic program uses $O(n^2)$ arithmetic operations after sorting.

From now on, we always assume $d\ge 2$.

\section{Transformation 1: phases}\label{sec:outer}

Archer, Levin, and Williamson~\cite[Corollary~8.4]{ALW2008}
give a deterministic strongly polynomial algorithm for weighted
metric MLP with positive integer customer weights and an
approximation ratio below $7.2$.

The same guarantee holds for positive rational weights.
Indeed, multiply all weights by the product of their denominators.
The resulting weights are integers with polynomial encoding length.
This multiplies every route cost by the same factor, so the
approximation ratio is unchanged.

Fix $0<\tau\le1$ and run this algorithm on the original instance.
Let $D_v>0$ be the completion times in the resulting customer
permutation. With $\beta=8$, we have
\begin{equation}\label{eq:baseline}
 \sum_v w_vD_v\le\beta\OPT.
\end{equation}

Set
\begin{equation}\label{eq:outer-parameters}
 \gamma=10/\tau,\qquad
 M=\left\lceil2\beta\gamma/\tau\right\rceil,\qquad
 R=2^M,\qquad \delta=R/\gamma.
\end{equation}
For each customer $v$, form the $M$ candidate thresholds
$D_v,D_v/2,\ldots,D_v/2^{M-1}$. Sort the list of all candidates for all customers in increasing
order. Keep the first one, then keep each candidate that is at least twice the last one kept. Number the retained thresholds from zero, and let $\mathcal T$ denote their set.

Fix a \emph{partition index} $s\in\{0,\ldots,M-1\}$ and let $\mathcal T_s$ be the
thresholds from $\mathcal T$ whose indices are congruent to $s$ modulo $M$. Assign each customer $v$ to the largest such threshold at most $D_v$. Each threshold $t$ used by partition $\mathcal T_s$ defines a phase with time interval \([t,Rt)\). We assign a customer $v$ to this phase if $t$ is the largest threshold from $\mathcal T_s$ at most $D_v$. The phase is nonempty if at least one customer is assigned to it.

For every candidate threshold $x$, the interval $(x/2,x]$ contains exactly one
retained threshold in $\mathcal T$. For $x=D_v/2^k$ with $0\le k<M$, these $M$ intervals partition
$(D_v/R,D_v]$. It therefore contains $M$ consecutive
thresholds in $\mathcal T$, exactly one for each $\mathcal T_s$. In particular, every customer $v$ is assigned to a phase with $t\le D_v<Rt$.

Number the nonempty phases in order of increasing time,
write $t_i$ for the lower threshold of phase $i$, and put
$L_i=\gamma t_i$. Denote its customer set by $V_i(s)$, its optimum
with delay $L_i$ by $O_i$, and the sum of these optima by $\Phi_s$.
Two consecutive nonempty phases satisfy $t_{i+1}\ge 2^M t_i=Rt_i$,
since their threshold indices differ by at least $M$. Thus their intervals do not overlap.

For phase $i$, restrict the initial approximate route to $V_i(s)$
by shortcutting. This gives a covering path $\mathcal H_i$ of length
less than $Rt_i$, since the last customer in this phase was served
before $Rt_i$. As $Rt_i=\delta L_i$, every phase is $\delta$-bounded.

\begin{lemma}\label[lemma]{lem:shift}
There is a partition index $s$ for which $\Phi_s\le(1+\tau)\OPT$.
\end{lemma}
\begin{proof}
Let $t_s(v)$ be the lower threshold of the phase containing $v$ for
partition index $s$. As $s$ varies, these thresholds include exactly
one from each interval $(D_v/2^{k+1},D_v/2^k]$, $0\le k<M$. Hence
\begin{equation}\label{eq:average-threshold}
 \frac1M\sum_{s=0}^{M-1} t_s(v)
 \le\frac{D_v}{M}\sum_{k=0}^{M-1}2^{-k}<\frac{2D_v}{M}.
\end{equation}

Now fix a phase partition. Let $C_v^*$ denote the completion times on an
optimal route for the original instance. Restrict that route to
$V_i(s)$ and start the restricted route at time $L_i$.
Its completion time for
$v\in V_i(s)$ is at most $C_v^*+L_i$. It follows that
\[
 O_i\le\sum_{v\in V_i(s)}w_v(C_v^*+L_i).
\]
The phase sets partition the customers, and $L_i=\gamma t_s(v)$
for customers of phase $i$. Summing the last inequality gives
\[
 \Phi_s\le\OPT+\gamma\sum_v w_vt_s(v).
\]
Average over the partition indices and use~\eqref{eq:average-threshold}
and~\eqref{eq:baseline}:
\[
 \frac1M\sum_s\Phi_s
 \le\OPT+\frac{2\gamma}{M}\sum_v w_vD_v
 \le\left(1+\frac{2\beta\gamma}{M}\right)\OPT
 \le(1+\tau)\OPT.
\]
At least one partition index has value no larger than this average.
\end{proof}

\begin{lemma}\label[lemma]{lem:assembly}
For a fixed phase partition, suppose a route of delayed cost $Z_i$
is available for each nonempty phase. These routes can be assembled
into a global route of cost at most $(1+\tau)\sum_i Z_i$. Phase $i$
starts at $r$ at time $L_i$ and returns by $\gamma Rt_i$.
\end{lemma}
\begin{proof}
Apply \cref{lem:complete} to $V_i(s)$ with covering path
$\mathcal H_i$, length bound $D=Rt_i$, and $\xi=\tau$.
The resulting closed tour has delayed cost at most $(1+\tau)Z_i$ and
returns by
\[
 L_i+5Rt_i/\tau
 =\gamma t_i+\tfrac{\gamma}{2}Rt_i
 \le\gamma Rt_i,
\]
since $R\ge2$. The next nonempty phase, if any, starts at
$L_{i+1}=\gamma t_{i+1}\ge\gamma Rt_i$. Execute these closed tours consequently,
waiting at the root when needed.
If a route happens to visit a customer
of a later phase, its actual completion time only decreases. Hence the
sum of the assigned phase costs is an upper bound on the global cost.
\end{proof}

Thus it remains to solve one $\delta$-bounded delayed instance.
If a local solver has approximation ratio $1+\alpha$, running it on every phase of every partition and keeping the best assembled route gives
\begin{equation}\label{eq:outer-guarantee}
 F(\text{output})\le(1+\tau)^2(1+\alpha)\OPT.
\end{equation}
Indeed, one factor $1+\tau$ comes from choosing a phase partition and the
other from assembling its phases.

\section{Transformation 2: a constant number of blocks}\label{sec:blocks}

Fix a nonempty $\delta$-bounded instance with delay $L$ and known covering path $\mathcal H$ of length at most $\delta L$, and fix
$0<\theta\le1/20$.

A \emph{block family} $\cP$ consists of endpoints
$z_0=r,z_1,\ldots,z_k\in V\cup\{r\}$ and paths $B_i$ from
$z_{i-1}$ to $z_i$. Every customer is assigned to exactly one block
that visits it, and each block has at least one assigned customer. Let
\[
 \ell_i=\len(B_i),\qquad
 q_i=\sum_{v\text{ assigned to }i}w_v,\qquad
 a_h=\sum_{i=h}^k q_i.
\]
Charge a customer at the
end of its assigned block, even if the route visits it earlier. The
resulting upper bound on the true delayed cost is
\begin{equation}\label{eq:surrogate}
 J_L(\cP)
 =\sum_{i=1}^k q_i\left(L+\sum_{h=1}^i\ell_h\right)
 =WL+\sum_{h=1}^k a_h\ell_h.
\end{equation}
The second equality follows because the length of block $h$ delays
all customers assigned to blocks $h,h+1,\ldots,k$.
This form of the objective is useful: once the assigned weights are
fixed, each block contributes only its length multiplied by a known coefficient.

Set
\begin{equation}\label{eq:block-bound}
 B=1+5\delta/\theta,\qquad
 K=\left\lceil\log_{1+\theta}B\right\rceil+1.
\end{equation}

\begin{lemma}[{Adapted from Sitters~\cite[Lemma~2.14]{Sitters2021}}]\label[lemma]{thm:blocks}
There is a family of at most $\min\{K,n\}$ blocks with
$J_L(\cP)\le(1+\theta)^2\OPT_L$ and
$\sum_i\ell_i\le5\delta L/\theta$.
\end{lemma}
\begin{proof}
Apply \cref{lem:complete} to an optimal
route, using $D=\delta L$ and $\xi=\theta$. This gives a route of cost at most
$(1+\theta)\OPT_L$ and length at most $5\delta L/\theta$,
whose delayed completion times lie in $[L,BL]$.

Round each such time $C_v$ up to the first value in
\[
 (1+\theta)L,\ (1+\theta)^2L,\ (1+\theta)^3L,\ldots
\]
that is at least $C_v$. Denote it by $\overline C_v$.
Then $\overline C_v\le(1+\theta)C_v$.
At most $K$ rounded values occur.

Group the customers by their rounded values. End a block at the last
visit in each nonempty group and assign that group to the block.
Its endpoint time is at most the common rounded value. The blocks
are consecutive pieces of the route up to its last visit,
so their total length is at most $5\delta L/\theta$. Here, a visit means the first visit to a customer. Each customer's block charge is at most
$w_v\overline C_v\le(1+\theta)w_vC_v$. Summing gives
\[
 J_L(\cP)\le(1+\theta)F_L(\text{obtained route})
 \le(1+\theta)^2\OPT_L.
\]
\end{proof}

\section{Transformation 3: portals and bounded crossings}
\label{sec:geometry}

Arora's PTAS for Euclidean TSP~\cite{Arora1998} divides space into
cells and restricts tours to cross their boundaries at designated
points, called portals, with a bounded number of crossings per face.
His structural estimate bounds the added length. To apply this
closed-tour estimate to our blocks, we add an auxiliary dimension.

Throughout this section, the input locations have integer coordinates
in $[0,N]^d$, where $N\ge1$ is a power of two. The delay satisfies
$L\ge1$. \cref{sec:rounding} obtains these requirements.

\subsection{The auxiliary dimension}

In $\R^{d+1}$, write a point as $(x,z)$ with $x\in\R^d$ and $z\in\R$, and use
the fixed norm
\begin{equation}\label{eq:lifted-norm}
 \|(x,z)\|_* = \|x\|_2+|z|.
\end{equation}
Let $\len_*$ denote length in this
norm. We call this auxiliary coordinate $z$ a \emph{height}.

Identify each input location $x$ with $(x,0)$, and write
$\Pi(x,z)=x$ for projection. For every polygonal walk $Q$,
\begin{equation}\label{eq:projection-length}
 \len_*(Q)=\len(\Pi Q)+\operatorname{TV}_z(Q),
\end{equation}
where $\operatorname{TV}_z(Q)$ is the sum of the absolute height
changes along its segments. This follows by summing
\eqref{eq:lifted-norm} over the segments.

Fix an integer height $H>0$ and put $x^+=(x,H)$.
A path $P$ from $s$ to $t$ of length $\ell$ gives the closed tour
\[
 s^+\longrightarrow(s,0)\overset{P}{\longrightarrow}(t,0)
 \longrightarrow t^+\longrightarrow s^+
\]
of length $\ell+2H+D$, where $D=\|s-t\|_2\le\ell$.
This also holds when $s=t$, with $D=0$.

\begin{lemma}\label[lemma]{lem:tour-extraction}
Let $S$ be a nonempty set of required points at height zero.
Put $D=\|s-t\|_2$.
Suppose a closed tour $C$ visits $S$, $s^+$, and $t^+$, and
$\len_*(C)<4H$. There is a subwalk $Q$ from $s^+$ to $t^+$
that visits all of $S$, has $\len_*(Q)\le\len_*(C)-D$, and satisfies
\begin{equation}\label{eq:extracted-length}
 \len(\Pi Q)\le\len_*(C)-D-2H.
\end{equation}
When $s=t$, take the whole tour $C$.
\end{lemma}
\begin{proof}
If $s=t$, take the whole tour starting at $s^+$. It reaches height
zero and returns to height $H$, so $\operatorname{TV}_z(C)\ge2H$.
The bound follows from~\eqref{eq:projection-length} with $D=0$.

Suppose $s\ne t$. Choose one occurrence of each upper point and
split the tour into the two arcs between them. An arc that reaches
height zero has vertical variation at least $2H$. If both arcs did
so, the tour would have length at least $4H$. Thus exactly one arc visits
height zero, and it contains every visit to $S$. Orient it from
$s^+$ to $t^+$ and call it $Q$. The other arc has length at least
$\|t^+-s^+\|_*=D$. Subtracting its length and then the vertical
variation of $Q$ gives~\eqref{eq:extracted-length}.
\end{proof}

For any path $Q$ with both endpoints at height $H$ and at
least one assigned customer at height zero, define
\begin{equation}\label{eq:corrected-length}
 c(Q)=\len_*(Q)-2H.
\end{equation}
Then $\len(\Pi Q)\le c(Q)$.

\paragraph{Lifted blocks.}
Let $S_i$ be the customer set assigned to block $B_i$, whose
endpoints are $z_{i-1}$ and $z_i$. A \emph{lifted block} for
$B_i$ is a polygonal walk $Q_i$ in $\R^{d+1}$ from
$z_{i-1}^+$ to $z_i^+$ that visits $(v,0)$ for every $v\in S_i$.
The projection $\Pi Q_i$ is a feasible block path in $\R^d$ with the
original endpoints and assigned customers. We call
$(Q_1,\ldots,Q_k)$ a \emph{lifted block family} when each $Q_i$
represents the corresponding block in this way.

For lifted blocks $Q_i$ from $z_{i-1}^+$ to
$z_i^+$ with the original customer assignments, define
\begin{equation}\label{eq:lifted-objective}
 \widehat J_L(Q_1,\ldots,Q_k)
 =WL+\sum_{i=1}^k a_i c(Q_i).
\end{equation}
Their projections join at $z_0,\ldots,z_k$. By~\eqref{eq:surrogate}
and~\eqref{eq:corrected-length}, their true delayed cost is at
most $\widehat J_L$.

\subsection{A common dissection}

Fix $0<\eta\le1$. \cref{thm:blocks} bounds the total length of
the blocks by $T=5\delta L/\theta$. Set
\begin{equation}\label{eq:height}
 H=\left\lceil2\max\{T,L\}\right\rceil,
 \qquad
 N_* = 2^{\lceil\log_2\max\{N,H\}\rceil}.
\end{equation}
The customers at height zero and the points at height $H$ above all
possible endpoints belong to $[0,N_*]^{d+1}$ and have integer coordinates.
Choose a \emph{dissection shift}
\[
 \alpha=(\alpha_1,\ldots,\alpha_{d+1})
 \in\{0,\ldots,N_*-1\}^{d+1}
\]
and start with the cube
\[
 \mathcal D_\alpha=\prod_{j=1}^{d+1}
 [-\alpha_j-\tfrac12,\,2N_*-\alpha_j-\tfrac12].
\]
Bisect every coordinate interval repeatedly until the side length
is one. The resulting cubes are the \emph{cells}, and the unit
cells are the \emph{leaves}. There are $O_d(N_*^{d+1})$ cells
and $O(\log N_*)$ levels. Every input or upper point is at
the centre of a leaf and lies on no cell boundary.

Let $p\ge2$ be a power of two. On each face of a cell of side
length $s$, place the regular grid of spacing $s/p$, including
its boundary. The vertices of this grid are called \emph{portals}. Each face has $m=(p+1)^d$ portals. These grids are
nested: a parent portal is a portal of every child face containing
it. A walk is \emph{portal respecting} if it crosses the boundaries of cells only at their portals. It is
\emph{$b$-light} if it crosses each face of each cell at most $b$
times, counting repeated crossings separately. All lifted blocks
use the same dissection, and the bound applies separately to each.

\paragraph{Boundary convention.}
The leaf labels determine the cell labels at higher levels. Distinguish copies of a boundary point in its incident leaves.
Subdivide segments at leaf boundaries and assign each resulting
piece lying on a boundary to one incident leaf, counting its length
there. A crossing is a zero-length transfer
between face-adjacent copies of the same point. At an edge or
corner, use a sequence of such transfers. Touching a boundary alone is
not a crossing. Geometric intersections do not join traversals
unless a connection is specified. Forgetting the cells gives
an ordinary walk of the same $\len_*$-length.

\subsection{The closed-tour estimate}

Arora bounds the expected added length in the proof of Theorem~5
in Section~2.3 of~\cite{Arora1998} and extends the argument to fixed
Minkowski norms in Section~2.4. For the interchange of sums over
dissection levels in that proof, we use the correction
in~\cite[footnote~7]{KNW2025}.

\begin{lemma}[Arora \cite{Arora1998}]\label[lemma]{lem:closed-tour-geometry}
For $0<\zeta\le1$, choose
\begin{equation}\label{eq:geom-parameters}
 b=O_d(\zeta^{-d}),\qquad
 p=O_d(\zeta^{-1}\log(N_*+2)),\qquad
 m=(p+1)^d.
\end{equation}
The constants can be chosen so that, for every fixed polygonal
closed tour $A$ with integer-coordinate vertices in
$[0,N_*]^{d+1}$, each shift $\alpha$ admits a portal-respecting,
$b$-light closed tour $C_\alpha$ visiting all vertices of $A$,
with
\[
 \len_*(C_\alpha)\le\len_*(A)+e_\alpha,
 \qquad e_\alpha\ge0,\qquad
 \E_\alpha[e_\alpha]\le\zeta\len_*(A).
\]
Here in the expectation the shift is uniform over its finite range.
\end{lemma}

The exponents in~\eqref{eq:geom-parameters} use ambient dimension
$d+1$. The norm~\eqref{eq:lifted-norm} is fixed and satisfies
$\|(x,z)\|_2\le\|(x,z)\|_*\le\sqrt2\|(x,z)\|_2$.
The cited expectation estimate therefore applies with constants depending only on $d$.
Applying it to an optimal tour on the vertices of $A$ gives the
stated bound for any $A$.

We now explain why Arora's expected-length argument applies to our dissection and to the norm $\|\cdot\|_*$. Arora's proof bounds the added length by charging it to crossings
of the unit grid. Fix a half-integer grid hyperplane and a dyadic cell size $s\le N_*$. This hyperplane
contains a cell face at scale $s$ for exactly one residue class
of the perpendicular shift modulo $s$. Since $s$ divides $N_*$,
this happens with probability $1/s$. Thus the probability that $s$
is the largest cell size with a face on this hyperplane is at most
$1/s$, as in Arora's proof~\cite[Section~2.3, Eq.~(5)]{Arora1998}.

Each segment $uv$ of the original tour crosses exactly
$\|u-v\|_1$ unit-grid hyperplanes. The total number of such crossings,
counted with multiplicity, is therefore at most
\[
 \sum_{uv\text{ in }A}\|u-v\|_1
 \le \sqrt d\,\len_*(A).
\]
These bounds demonstrate why we can use \cref{lem:closed-tour-geometry}.

\subsection{One height and one shift for all blocks}

\begin{lemma}\label[lemma]{lem:geometric}
Let $B_1,\ldots,B_k$ be a block family $\cP$ with $k\le K$,
$\ell_i=\len(B_i)$, and $\sum_i\ell_i\le T$.
Use $T$, $H$, $N_*$, and $\eta$ as above, and set
\begin{equation}\label{eq:tour-accuracy}
 \zeta=\frac{\eta L}{(2K+1)H}.
\end{equation}
With the portal parameters of~\eqref{eq:geom-parameters}, there
exist a common dissection shift $\alpha$ and portal-respecting,
$b$-light lifted blocks $Q_i$ from $z_{i-1}^+$ to $z_i^+$, with
the original customer assignments, such that
\[
 \widehat J_L(Q_1,\ldots,Q_k)\le(1+\eta)J_L(\cP),
 \qquad \sum_{i=1}^k c(Q_i)\le T+\eta L.
\]
\end{lemma}
\begin{proof}
Shortcut each block to its prescribed start, its assigned customers
in their order of first visit, and its prescribed end. Its vertices
are then integer points and its length does not increase. Keep
$\ell_i$ as an upper bound on that length.
Close it using the upper endpoints as above. Write $A_i$ for
this closed tour and $D_i=\|z_i-z_{i-1}\|_2$. Then
\[
 \len_*(A_i)\le\lambda_i:=\ell_i+2H+D_i\le3H,
 \qquad
 \sum_i\lambda_i\le2T+2kH\le(2K+1)H.
\]
Here $D_i\le\ell_i$ and $T\le H/2$.

Apply \cref{lem:closed-tour-geometry} to all $A_i$ using
the same random shift. Let $e_i\ge0$ be their added-length
bounds. Linearity of expectation gives
\[
 \E_\alpha\sum_i e_i
 \le\zeta\sum_i\lambda_i\le\eta L.
\]
Choose a shift $\alpha$ for which $\sum_i e_i\le\eta L$.
Each resulting tour $C_i$ satisfies
\[
 \len_*(C_i)\le3H+\eta L\le\tfrac72H<4H.
\]
\cref{lem:tour-extraction} provides the required path $Q_i$,
or the whole tour when its upper endpoints coincide, with
\[
 c(Q_i)\le\len_*(C_i)-D_i-2H\le\ell_i+e_i.
\]
Each $Q_i$ retains the portals and crossing bound of its tour,
because it is a subwalk whose endpoints are interior input points.
Summing the bounds on $c(Q_i)$ gives the claimed total length bound.
Since $a_i\le W$,
\begin{align*}
 \widehat J_L(Q_1,\ldots,Q_k)
 &\le J_L(\cP)+\sum_i a_i e_i\\
 &\le J_L(\cP)+W\sum_i e_i\\
 &\le J_L(\cP)+\eta WL
 \le(1+\eta)J_L(\cP).
\end{align*}
\end{proof}

Apply \cref{thm:blocks,lem:geometric} with
$\eta=\theta$. For at least one dissection shift there is a lifted
family with endpoints above customers or the root satisfying
\begin{equation}\label{eq:restricted-witness}
 \widehat J_L(Q_1,\ldots,Q_k)\le(1+\theta)^3\OPT_L,
 \qquad \sum_i c(Q_i)\le T+\eta L.
\end{equation}
The algorithm computes $H$ from $T$ and tries all shifts. Since
$L\ge1$, we have $H/L\le2\max\{T/L,1\}+1$. Thus $\zeta^{-1}$ depends
only on the accuracy parameters and $\delta$.

The dynamic program searches lifted paths with at least one assigned
customer each. Projection preserves the
cost bound~\eqref{eq:lifted-objective}, but need not preserve the
crossing restrictions, so it is applied only after the lifted
paths have been constructed.

\section{Rounding locations in a phase}\label{sec:rounding}

We put the customers on the integer grid required in
\cref{sec:geometry}. Fix a phase with delay $L>0$, $n_0$
positive-weight customers of total weight $W$, and a known covering
path of length at most $\delta L$. Translate the root to the origin.
Set
\[
 \rho=\frac{\theta L}{40n_0},\qquad
 E_0=2n_0\rho W=\frac{\theta WL}{20}.
\]
Round each coordinate to the nearest multiple of $\rho/d$, keeping
the root fixed. Each customer moves by at most $\rho$ in Euclidean
distance. Denote the original and rounded instances by $I$ and
$\widetilde I$.

\begin{lemma}\label[lemma]{lem:coordinates}
For every permutation $\pi$ of the original customer identities,
\[
 \bigl|F_L^I(\pi)-F_L^{\widetilde I}(\pi)\bigr|\le E_0.
\]
The rounded instance has a known covering path of length at most
$(\delta+1)L$.
\end{lemma}
\begin{proof}
Moving the endpoints of an edge changes its length by at most $2\rho$.
A completion time sums at most $n_0$ edge lengths, so its change is at
most $2n_0\rho$. Multiplying by the weights and summing gives $E_0$.
Rounding the locations adds at most $2n_0\rho\le L$ to the supplied covering path length.
\end{proof}

Merge customers at the same rounded location by summing their weights.
This preserves the optimum of the rounded instance. A permutation of the
merged locations expands to a permutation of the original identities
by listing customers at each location consecutively. Customers at the
rounded root can be served at time $L$.

Divide coordinates by $\rho/d$ to make them integer.
Then subtract the coordinatewise minimum of the root and all rounded
customer locations from every point.
Since the original customers are within $\delta L$ of the root,
the scaled locations fit in $[0,N]^d$ for a power of two
\begin{equation}\label{eq:grid-side}
 N=O_d(1+\delta n_0/\theta).
\end{equation}
Scale the delay and all lengths by the same factor $d/\rho$.
This multiplies each cost by $d/\rho$. The scaled delay equals
$40dn_0/\theta\ge1$. We write $L$ for the delay in the current units.

The merged instance has at most $n_0$ customers and total weight $W$.
Its covering path has length at most $\Delta L$, where
$\Delta=\delta+1$. Thus \cref{sec:blocks,sec:geometry}
apply with
\begin{equation}\label{eq:local-parameters}
 \Delta=\delta+1,\qquad
 T=5\Delta L/\theta,\qquad
 K=\left\lceil\log_{1+\theta}(1+5\Delta/\theta)\right\rceil+1,
 \qquad \eta=\theta .
\end{equation}
In particular, the height and the size of the dissection in
\eqref{eq:height} satisfy
\begin{equation}\label{eq:lifted-grid-bounds}
 H,N_*=O_{d,\delta,\theta}(n_0),\qquad
 H/L=O(\Delta/\theta),\qquad \zeta^{-1}=O(K\Delta/\theta^2).
\end{equation}

\section{Computing the restricted routes}\label{sec:dp}

Work with the rounded and merged instance of \cref{sec:rounding}, with parameters from \eqref{eq:local-parameters}.
Let $n\le n_0$ be the number of distinct customer locations. Customers
lie at height zero. Any location $x$ of a customer or the root is
eligible as a block endpoint, with upper point $x^+=(x,H)$.

\subsection{Rounding the weight sums}

For a cell $C$, let $W_C$ be the total weight of its customers. We
can compute $W_C$ before assigning any customers to blocks. The
rounding step for this cell will be $W_C/m_w$, where the same
positive integer $m_w$ is used throughout the dissection.

A leaf has level zero and the outermost cell has level
$h_*=\log_2(2N_*)$. Set
\begin{equation}\label{eq:weight-grid}
 T_0=\max\{L,T+\eta L\},\qquad
 m_w=\left\lceil\frac{K(h_*+1)T_0}{\eta L}\right\rceil.
\end{equation}
Here $T_0$ bounds the sum of the corrected block lengths in
\eqref{eq:restricted-witness}. If $W_C>0$, let $\mathcal R_C(x)$ be
the smallest multiple of $W_C/m_w$ at least $x$. In every cell set
$\mathcal R_C(0)=0$. This is the only value needed when $W_C=0$.

For a fixed assignment, write $q_i(C)$ for the true weight of the
customers in $C$ assigned to block $i$. At a leaf, set
$\widehat q_i(C)=\mathcal R_C(q_i(C))$. At a parent, sum the rounded
values of its children and round that sum using the parent's step.

\begin{lemma}\label[lemma]{lem:weight-grid}
In a cell $C$ of level $h$,
\[
 q_i(C)\le\widehat q_i(C)
 \le q_i(C)+\frac{(h+1)W_C}{m_w}.
\]
The rounded value is positive exactly when the cell contains a
customer assigned to the block. For $W_C>0$ its integer index
$\widehat q_i(C)/(W_C/m_w)$ lies between $0$ and $m_w+h+1$.
\end{lemma}
\begin{proof}
Rounding up adds at most $W_C/m_w$ and preserves zero. At a parent
of level $h$, the errors of its children sum to at most
\[
h\sum_{C'\text{ child of }C}W_{C'}/m_w=hW_C/m_w.
\]
Rounding
once more proves the bound by induction. All customer weights are
positive. The bound on the index follows from $q_i(C)\le W_C$.
\end{proof}

\subsection{Enumeration and objective}

Enumerate the number of blocks $1\le k\le\min\{K,n\}$ and the
dissection shift $\alpha$. Let $C_*$ be the outermost cell, so that
$W_{C_*}=W$. Enumerate also the positive rounded weights
$\bar q_1,\ldots,\bar q_k$ in $C_*$. Each is a multiple of $W/m_w$ with
index at most $m_w+h_*+1$. Keep only vectors satisfying
\begin{equation}\label{eq:weight-caps}
 \sum_{i=1}^k \bar q_i\le (1+\eta)W .
\end{equation}
Every assignment to nonempty blocks has such a vector. Indeed,
\cref{lem:weight-grid} bounds the sum of the errors in $C_*$ by
$k(h_*+1)W/m_w\le\eta W$, using $T_0\ge L$. For a given vector the outermost cell $C_*$ must realize $\widehat q_i(C_*)=\bar q_i$ for every block.

Write $\bar a_i=\sum_{j=i}^k\bar q_j$. For this run the dynamic
program minimizes $\sum_i\bar a_i\len_*(Q_i)$. To compare solutions
from different runs, use
\begin{equation}\label{eq:dp-objective}
 \overline J_L
 =WL+\sum_{i=1}^k\bar a_i c(Q_i)
 =WL+\sum_{i=1}^k\bar a_i\len_*(Q_i)-2H\sum_{i=1}^k\bar a_i .
\end{equation}
The subtracted term can change with the weights $\bar q_1,\ldots,\bar q_k$.

\begin{lemma}\label[lemma]{lem:rounded-objective}
For any lifted block family whose customer assignment gives the chosen
rounded weight vector in $C_*$, $\overline J_L$ is an upper bound on
the true delayed cost after projection. For the family in
\eqref{eq:restricted-witness}, its rounded root weights also satisfy
\[
 \overline J_L\le\widehat J_L+\eta WL .
\]
\end{lemma}
\begin{proof}
Let $q_i$ be the actual weight of block $i$ and
$a_i=\sum_{j=i}^k q_j$. \cref{lem:weight-grid} gives
\[
 0\le \bar a_i-a_i\le \frac{k(h_*+1)W}{m_w}.
\]
Each block has positive rounded root weight, hence contains a
customer at height zero. Its corrected length $c(Q_i)$ is
nonnegative and bounds the length of its projection. It follows that
$\overline J_L\ge\widehat J_L$, which bounds the projected cost
by~\eqref{eq:lifted-objective}. For the family in
\eqref{eq:restricted-witness}, we also know that $\sum_i c(Q_i)\le T_0$.
Consequently,
\[
 \overline J_L-\widehat J_L
 =\sum_i(\bar a_i-a_i)c(Q_i)
 \le\frac{k(h_*+1)W}{m_w}T_0\le\eta WL .
\]
\end{proof}

\subsection{States and transitions}

The dynamic program chooses where consecutive blocks meet. Give the
unknown upper points $z_1^+,\ldots,z_k^+$ labels $1,\ldots,k$.
Placing label $j<k$ at $x^+$ makes that point both the end of block
$j$ and the start of block $j+1$. Label $k$ gives the final endpoint.
The start of block $1$ is fixed at $r^+$. Several labels may be placed
at the same upper point. The two ends of each block are distinct
occurrences even when they have the same location.

A state of a cell $C$ records the set $I_C\subseteq\{1,\ldots,k\}$
of labels placed in $C$. For each block it also records
\begin{enumerate}[leftmargin=*,itemsep=2pt]
 \item the boundary crossing occurrences, each specified by its face,
       its portal, and the leaf inside $C$ containing the copy of the
       portal used at that occurrence, with at most $b$ crossings per face,
 \item a pairing of those crossings and the ends of the block in $C$.
       Each pair gives the ends of one piece of the path inside $C$,
 \item the rounded assigned weight $\widehat q_i(C)$, stored as         the integer $j_i$ such that $\widehat q_i(C)=j_i W_C/m_w$,
    with $j_i=0$ when $W_C=0$.
\end{enumerate}
The labels in $I_C$, together with the known position of $r^+$,
determine which block ends the pairing must include. In particular,
an internal label contributes one end to each of two successive
blocks. The state of a cell need not specify the leaf that contains
a label. The choice is recovered from its children.

Each table entry stores the minimum sum of piece lengths weighted
by their block coefficients $\bar a_i$, or $+\infty$ if the state
is infeasible. Each customer in the cell must be assigned to exactly
one block that visits it. We retain choices attaining finite entries
for reconstruction.

\paragraph{Leaves.}
At a leaf centred at an eligible upper point $x^+$, try every subset
of labels $I_C$, all placed at that point. At every other leaf require
$I_C=\varnothing$. Include the fixed start in the leaf containing
$r^+$, regardless of $I_C$.

A leaf contains at most one customer, at its centre. For each
pairing, join the ends of every pair by a straight segment. If the
leaf contains a customer, try each paired piece in turn to serve it.
Replace that piece by the two segments through the customer and
assign its whole weight to the corresponding block. Its rounded
weight is exactly $W_C$, which is $m_w$ steps on the cell grid.
All other blocks get weight zero. Discard a pairing that cannot
serve the customer. These choices give the exact leaf minimum for
exact segment lengths. Any piece can be straightened between its
ends, or on either side of its assigned visit, and convexity keeps
the segments inside the leaf.

\paragraph{Parent cells.}
Try each combination of states of the $2^{d+1}$ children. Their
label sets must be pairwise disjoint and have union $I_C$. An
occurrence on an internal face must match an occurrence of the same
block at the same portal in the adjacent child. The two recorded leaves
must share a unit face containing that portal. Discard a combination
if some internal occurrence cannot be matched. When several occurrences
coincide, try all matchings satisfying this condition. Follow the
connections to obtain the parent pairing. Every remaining crossing
keeps its recorded leaf and must be at a parent portal, with at most
$b$ crossings per block and face. Edges and corners use the boundary
convention of \cref{sec:geometry}.

Discard a closed component without a block endpoint. If one piece
already joins both ends of a block, discard any combination with
another piece of that block in the same cell. For each block, sum the
children's rounded \emph{weights}, converting their stored indices
using their own steps. Round the sum using $\mathcal R_C$. The cost
of the combination is the sum of the child costs.

At the outermost cell require $I_{C_*}=\{1,\ldots,k\}$, no boundary
crossings, and one path joining the two ends of each block. Its
rounded weight must equal the enumerated $\bar q_i>0$.

\begin{lemma}\label[lemma]{lem:dp}
For fixed $k$, $\alpha$, and rounded weights in $C_*$, the dynamic program
with exact lengths minimizes $\sum_i\bar a_i\len_*(Q_i)$ over all
portal-respecting, $b$-light lifted block families realizing those
weights, including all allowed choices of block endpoints.
\end{lemma}
\begin{proof}
At a leaf the allowed centres, pairings, and assigned visits are
enumerated explicitly. Cutting a feasible family at the child
boundaries gives one of the combinations considered at its parent.
Each endpoint label lies in exactly one child because its upper
point lies in the interior of a leaf. Conversely, a retained
combination puts each label in one place and joins precisely the
path pieces specified by the parent pairing. Replacing a child
solution by a cheaper one with the same state preserves all outside
connections because the face, portal, and incident leaf of each crossing
are fixed. It also preserves each rounded child weight, so every
ancestor computes the same rounded weights. Induction gives the
minimum cost in every state. The requirements at the outermost cell
give exactly the families in the statement.
\end{proof}

\subsection{Running time}

A cell has $2(d+1)$ faces. Each block has at most $b$ crossings
per face and two further ends, so a state has at most
$2(d+1)kb+2k$ ends. The number of ways to pair and label them
depends only on $d,k,b$. There are $2^k$ choices for $I_C$.
Each boundary end has at most $m$ portal choices on its face. For a
fixed face and portal there are at most $2^d$ incident leaves inside
the cell, since each of the $d$ coordinates along the face gives at
most two choices. Each rounded weight has at most $g=m_w+h_*+2$
possible indices.
The number $S$ of states per cell therefore satisfies
\begin{equation}\label{eq:states}
 S\le f_d(K,b)\, g^k m^{O_d(kb)} .
\end{equation}
Here $f_d(K,b)$ includes the choices of endpoint labels. In
particular, the number of possible locations of a label does not
appear as a factor for each endpoint.

For each parent we try at most $S^{2^{d+1}}$ child combinations.
The number of matchings per combination is bounded in terms of
$d,K,b$. There are $O_d(N_*^{d+1})$ cells for each of $N_*^{d+1}$
shifts. Each block count gives at most $g^k$ root weight vectors.
Including the search for grid indices and the evaluation of leaf
lengths, the local computation takes at most
\begin{equation}\label{eq:dp-time}
 f_d(K,b)\,N_*^{2(d+1)}
     g^{(1+2^{d+1})K+2}m^{O_d(Kb)}
\end{equation}
operations, apart from a fixed polynomial for preparation. At a
parent an upward rounding can be found by searching its $g$ indices.
The rational lengths used at leaves can also be found by comparisons,
as shown in \cref{app:arithmetic}.

For fixed $d,\delta,\theta$, we have $N_*=O_{d,\delta,\theta}(n_0)$
and $T_0/L$ is bounded in terms of $\delta,\theta$. By
\eqref{eq:weight-grid} and~\eqref{eq:geom-parameters},
\[
 g=O_{d,\delta,\theta}(\log(n_0+2)),\qquad
 m=O_{d,\delta,\theta}((\log(n_0+2))^d),\qquad
 b=O_d\bigl((K\Delta/\theta^2)^d\bigr).
\]
Thus~\eqref{eq:dp-time} is bounded by
$C(n_0+2)^{2(d+1)}(\log(n_0+2))^A$, where $C,A$ depend only on
$d,\delta,\theta$. For $A>0$ and $x\ge1$,
\begin{equation}\label{eq:absorb-log}
 (\log x)^A\le (A/e)^A x ,
\end{equation}
as follows by maximizing $y^Ae^{-y}$ with $y=\log x$. For $A=0$
the logarithmic factor is one. The exponent of $n_0$ in the local
running time therefore depends only on $d$.

\subsection{Returning to the original phase}

Use the rational upper segment lengths from \cref{app:arithmetic}
in the tables. Over all runs, choose the reconstructed family of
smallest corrected objective~\eqref{eq:dp-objective}. Project its
paths and concatenate them in block order. Read off the assigned
customer visits along these paths and shortcut to this order.
Any customer at the rounded root can be placed first. Expand each
merged location into its original customers and traverse the resulting
order at the original coordinates.

\begin{theorem}\label{thm:local}
For fixed $d,\delta$ and $0<\theta\le1/20$, the algorithm returns
a route of delayed cost at most $(1+5\theta)\OPT_L$ using
$f_{d,\delta}(\theta)(n_0+1)^{O_d(1)}$ arithmetic operations.
Here $n_0$ is the phase size before merging.
\end{theorem}
\begin{proof}
Express all costs in the original distance units. Let $O=\OPT_L$
be the original phase optimum and $\widetilde O$ the optimum after
rounding the coordinates and merging equal locations.
\cref{lem:coordinates} gives $\widetilde O\le O+E_0$, where
$E_0=\theta WL/20$.

Apply~\eqref{eq:restricted-witness} and
\cref{lem:rounded-objective} to the merged instance and rescale
to the original units. They give a lifted family whose rounded
objective is at most $(1+\theta)^3\widetilde O+\theta WL$.
The shift, weights, and endpoint labels of this family are all among
the choices of the dynamic program. By \cref{lem:dp} and
\cref{app:arithmetic}, using rational upper segment lengths
adds at most $E_0$ to the minimum over those choices.

Projection and shortcutting give a route of no greater cost for the
rounded locations. Restoring the original coordinates adds at most
$E_0$ by \cref{lem:coordinates}. Hence
\[
 F_L(\text{output})
 \le (1+\theta)^3 O+\theta WL+\bigl((1+\theta)^3+2\bigr)E_0 .
\]
For $0<\theta\le1/20$, we have $(1+\theta)^3\le1+7\theta/2\le2$.
Since $WL\le O$, the right-hand side is at most
$(1+7\theta/2+\theta+4\theta/20)O\le(1+5\theta)O$.
The running time follows from~\eqref{eq:dp-time} and~\eqref{eq:absorb-log}.
\end{proof}

\section{The complete scheme}\label{sec:complete}

\begin{proof}[Proof of \cref{thm:main}]
Replace $\eps$ by $\min\{\eps,1\}$, and choose
\begin{equation}\label{eq:final-parameters}
 \tau=\eps/30,\qquad \theta=\tau/5,\qquad
 \kappa=\min\{\tau,1/1000\}.
\end{equation}
Serve the customers at the root and set aside zero-weight customers.
Compute the rational metric $\bar d$ from \cref{app:arithmetic}
and use it for the initial approximation and phase thresholds.
It satisfies $d\le\bar d\le(1+\kappa)d$ and preserves
\eqref{eq:baseline}. For every nonempty phase of every phase partition,
run the Euclidean local algorithm of \cref{sec:dp}.
Complete and assemble the resulting orders in the metric $\bar d$
using \cref{lem:assembly}. Choose the assembled route of
smallest $\bar d$-cost, remove its waits, and traverse its customer
order in Euclidean space. Finally, append any unvisited zero-weight
customers.

For the partition index supplied by \cref{lem:shift}, let $Z_i$ be the
computed Euclidean delayed costs and $O_i$ the Euclidean phase
optima. The $\bar d$-cost of each local order is at most
$(1+\kappa)Z_i$. The two phase lemmas and
\cref{thm:local} therefore give
\begin{align*}
 F(\text{output})
 &\le(1+\tau)(1+\kappa)\sum_i Z_i\\
 &\le(1+\tau)(1+\kappa)(1+5\theta)\sum_i O_i\\
 &\le(1+\tau)^2(1+\kappa)(1+5\theta)\OPT\\
 &\le(1+\eps/30)^4\OPT\le(1+\eps)\OPT .
\end{align*}
The last inequality follows from $(1+x)^4\le1+5x$ for
$0\le x\le1/30$.

All partition indices, dissection shifts, and discrete choices are enumerated, and the initial
approximation is deterministic. Each local call has polynomial
exponent depending only on $d$, and there are at most $Mn$ calls.
This proves the theorem.
\end{proof}

\paragraph{Running time of the complete scheme.}
Let $n$ here denote the total number of customers in the original
instance. For the parameters in~\eqref{eq:final-parameters},
\[
 M=O(\eps^{-2}),\qquad
 \log(\delta+1)=O(\eps^{-2}),\qquad
 K=O(\eps^{-3}).
\]
The last bound follows from~\eqref{eq:local-parameters} and
$\log(1+\theta)=\Theta(\theta)$. The crossing bound $b$ depends
only on $d,\eps$. The number $g$ of weight indices per block and
the number $m$ of portals per face are bounded by powers of
$\log(n+2)$ with factors depending only on $d,\eps$. Substituting
these estimates into~\eqref{eq:dp-time}, applying~\eqref{eq:absorb-log},
and including all $Mn$ local calls gives
\begin{equation}\label{eq:total-time}
 f_d(\eps)(n+1)^{O_d(1)}
\end{equation}
arithmetic operations and comparisons, counting each Euclidean
norm evaluation as one operation. The exponent depends on $d$
but not on $\eps$.
\cref{app:arithmetic} gives the implementation with
bit complexity $f_d(\eps)(n+\mathcal B+1)^{O_d(1)}$.

\section{Nonexistence of an FPTAS}\label{sec:no-fptas}

An FPTAS has running time polynomial in the binary input length
and $1/\eps$. We record an elementary consequence of the
NP-completeness of Hamilton path in grid graphs. The argument
gives an explicit gap and requires no exact comparison of sums
of square roots.

\begin{theorem}\label{thm:no-fptas}
For every fixed $d\ge2$, weighted Euclidean MLP in
$\R^d$ has no FPTAS unless $\mathrm P=\mathrm{NP}$. This holds
even when all customer weights are one and the root and customers
occupy distinct points with integer coordinates.
\end{theorem}
\begin{proof}
A grid graph is a finite vertex-induced subgraph of the integer
lattice: two vertices are adjacent exactly when their Euclidean
distance is one. Deciding whether such a graph has a Hamilton
path, with neither endpoint prescribed, is NP-complete
\cite{IPS1982}.

Let $G$ be a grid graph on $N\ge2$ vertices. For each vertex $r$,
form an instance $I_r$ of planar Euclidean MLP with root $r$ and
a unit-weight customer at every other vertex. Write $n=N-1$ and
set
\[
 B_n=\sum_{j=1}^n j=\frac{n(n+1)}2.
\]
As in \cref{sec:model}, shortcutting allows us to use
customer permutations. For any such permutation, with $\pi_0=r$,
\[
 F(\pi)=\sum_{h=1}^n (n-h+1)\|\pi_h-\pi_{h-1}\|_2
 \ge B_n,
\]
because distinct integer points have distance at least one.
Equality holds exactly when every consecutive pair is adjacent
in $G$, so the permutation is a Hamilton path starting at $r$.
If the permutation is not such a path, one of its consecutive
pairs has distance at least $\sqrt2$. Its coefficient in the sum
is at least one, and hence
\begin{equation}\label{eq:fptas-gap}
 F(\pi)\ge B_n+\sqrt2-1.
\end{equation}

Suppose an FPTAS exists. Run it on $I_r$ for every vertex $r$,
using
\[
 \eps=\frac{1}{4B_n}.
\]
If $G$ has a Hamilton path, choose $r$ to be one of its endpoints.
Then $\OPT(I_r)=B_n$, and the returned permutation satisfies
\[
 F(\pi)\le(1+\eps)B_n=B_n+\frac14
 <B_n+\sqrt2-1.
\]
By~\eqref{eq:fptas-gap}, this permutation must itself be a
Hamilton path in $G$. Conversely, if $G$ has no Hamilton path,
none of the returned permutations can be one.

We can therefore decide Hamilton path by checking whether any
returned permutation uses only edges of $G$. These checks use
only integer coordinates, with no need to evaluate the latency
costs. There are $N$ calls and $1/\eps=4B_n=O(N^2)$, so the total
running time is polynomial in the input length. This implies
$\mathrm P=\mathrm{NP}$. Finally, appending zero coordinates
embeds every constructed instance isometrically into $\R^d$
for any fixed $d\ge2$.
\end{proof}

\section*{Use of AI tools}
The authors used ChatGPT (OpenAI) to assist in developing and refining the mathematical proofs, as well as for language editing. The authors reviewed the resulting material and take full responsibility for the content of the paper.

\appendix
\section{Rational arithmetic}\label[appendix]{app:arithmetic}

We give a rational implementation of the local tables and the outer
phase construction. For $0<\eps\le1$ we may use a dyadic accuracy in
$(\eps/2,\eps]$, found by repeated halving. Thus the precision of the
accuracy parameters can be bounded in terms of $\eps$.

\paragraph{Local tables.}
Each leaf has at most $k((d+1)b+1)$ pairs of ends. Passing a piece
through the leaf's customer adds at most one more segment. Therefore
any reconstructed family uses at most
\[
 G=(2N_*)^{d+1}\bigl(K((d+1)b+1)+1\bigr)
\]
segments. Round each segment's $\|\cdot\|_*$-length up to a
multiple of
\[
 \omega=\frac{\theta L}{20(1+\eta)G}.
\]
Keep $WL$ and the correction $2H\sum_i\bar a_i$ exact. By
\eqref{eq:weight-caps}, every coefficient $\bar a_i$ is at most
$(1+\eta)W$. The rounded objective of a family in leaf form thus
lies between its exact objective and that value plus
$(1+\eta)WG\omega=\theta WL/20$. An exact optimum for a run can
be put in leaf form by straightening its pieces, as in
\cref{lem:dp}. The same error bound holds for every run, so
minimizing over runs adds at most $\theta WL/20$. In the original
distance units this is $E_0$ in \cref{thm:local}.

For a segment with displacement $(x,z)$, test whether
$\|x\|_2+|z|\le j\omega$ by first checking $j\omega\ge |z|$ and
then comparing $\sum_\ell x_\ell^2$ with $(j\omega-|z|)^2$.
In scaled units a segment has length at most $2(\sqrt d+1)N_*$,
and $N_*/\omega$ is polynomial in $n_0$ for fixed $d,\delta,\theta$.
Binary search thus finds the least admissible $j$ in
$O_{d,\delta,\theta}(\log(n_0+2))$ comparisons, including equality
and zero-length segments.

The cell weights also have short exact representations. Let $D_w$
be the product of the denominators of the input weights. Every
$W_C$ is a multiple of $1/D_w$. Every rounded value is an integer
index times $W_C/m_w$, so it is a multiple of $1/(D_wm_w)$, at
every depth. Child weights can therefore be added and compared with
the parent grid exactly. The coefficients $\bar a_i$ share the
same denominator. Table costs are sums of at most $G$ products of
such a coefficient with a multiple of $\omega$.

\paragraph{Phase thresholds and assembly.}
Use $\kappa$ from~\eqref{eq:final-parameters}. For distinct
locations $u,v$, put $a=\|u-v\|_\infty>0$. The ratio
$d(u,v)/a$ is in $[1,d]$. Bisect this interval, using comparisons
of rational squares, until its width is at most $\kappa$, and
multiply the upper endpoint by $a$ to obtain $e(u,v)$. This takes
$O_d(1+\log(1/\kappa))$ arithmetic operations per pair. Since the
ratio is at least one,
\[
 d(u,v)\le e(u,v)\le(1+\kappa)d(u,v).
\]
Set $e(u,v)=0$ for coincident locations. Let $\bar d$ be the
shortest-path metric in the complete graph with edge lengths $e$.
The triangle inequality for $d$ gives
\[
 d(u,v)\le\bar d(u,v)\le(1+\kappa)d(u,v).
\]
Coincident locations can be merged for the initial approximation
and expanded afterwards. To implement the thresholds of Archer, Levin,
and Williamson, divide all $\bar d$-distances in each of their
preprocessed blocks by the distance from the root to the nearest
remaining customer in that block. Set $q=401/400$, $P=512$, and $A=q^P$. For each
$k=0,\ldots,P-1$, use the grid $\{q^k A^i:i\in\mathbb Z\}$ within
their search interval and keep the best order over all $P$ choices.
For any $x>0$, the first grid points at or above $x$, one from each
grid, are $u,uq,\ldots,uq^{P-1}$ in increasing order, with
$x\le u<qx$. Their average is at most $(qx/P)\sum_{j=0}^{P-1}q^j$.
The concatenation bound in~\cite[Section~7]{ALW2008}, together with
the weighted analysis in the proof of Theorem~8.3 of the same paper,
gives an approximation ratio of at most
\[
 2\,\frac{A+1}{A-1}\,\frac{q}{P}\sum_{j=0}^{P-1}q^j
 =\frac{2(A+1)q}{P(q-1)}<7.192 .
\]
The range bound in that proof gives $O(n\log(n+1))$ successive
multiplications by $A$ for each offset in the scaled units, so all
thresholds have polynomial bit length.

Use the completion times of the resulting order in the original
units of $\bar d$ as $D_v$. Then
\[
 \sum_v w_vD_v\le7.2(1+\kappa)\OPT<8\OPT,
\]
so~\eqref{eq:baseline} still holds. \cref{lem:shift} uses the
times $D_v$ only to define the phases and bound their delays.
Its comparison with the optimal route therefore holds for the
Euclidean phase optima.
The covering path $\mathcal H_i$ has $\bar d$-length less than
$Rt_i$, and its Euclidean length is at most its $\bar d$-length.
Since each $D_v$ is rational, so are the phase thresholds, delays, and
coordinate-rounding scales.

The local solver still uses Euclidean geometry. If its returned
order has Euclidean delayed cost $Z_i$, its delayed cost measured
using $\bar d$ is at most $(1+\kappa)Z_i$. Perform the completion
operation and assembly using $\bar d$, and compare the assembled
routes by their rational costs in this metric. \cref{lem:assembly}
then bounds the selected cost by
$(1+\tau)(1+\kappa)\sum_i Z_i$ for the good phase partition.
Traversing each resulting edge in Euclidean space takes no more
time. Since $\kappa\le\tau$, this accounts for the one additional
factor used in the proof of \cref{thm:main}.

Let $\mathcal B$ be the binary encoding length of the input. The
initial metric approximations can use the same number of bisections
for every pair. After clearing the coordinate denominators, their
only additional denominator is a power of two with
$O_d(1+\log(1/\kappa))$ bits. Shortest-path distances and baseline
completion times are sums of polynomially many such lengths.
Dividing $D_v$ by $2^k$ adds at most $k<M$ bits to its denominator.
The construction in \cref{sec:outer}
generates these levels in a number of operations independent of
the bit lengths of the coordinates.

The scaled local coordinates, portal denominators, $m_w$, and $G$
are bounded by $f_d(\eps)(n+1)^{O_d(1)}$. In~\eqref{eq:local-parameters},
$K$ can be found by successive multiplications by $1+\theta$.
Doubling gives the required powers of two, and coordinate rounding
can use binary search over the integer grid in~\eqref{eq:grid-side}.
The product $D_w$ has at most the sum of the bit lengths of the input
denominators. Together with the common denominators for the local
tables, these bounds give encoding length
$f_d(\eps)(\mathcal B+\log(n+2)+1)^{O_d(1)}$ for all intermediate
numbers. Rational arithmetic therefore gives total bit complexity
$f_d(\eps)(n+\mathcal B+1)^{O_d(1)}$.

\end{document}